\documentclass[12pt]{article}
\usepackage{amssymb,amsmath,amsfonts}

\def\ba{\begin{array}}
\def\ea{\end{array}}

\def\ben{\begin{enumerate}}
\def\een{\end{enumerate}}
\def\beqan{\begin{eqnarray*}}
\def\eeqan{\end{eqnarray*}}
\def\btab{\begin{tabular}}
\def\etab{\end{tabular}}
\def\bit{\begin{itemize}}
\def\eit{\end{itemize}}

\newcommand{\lra}{\longrightarrow}

\newcommand{\mbb}[1]{\mathbb{#1}}
\newcommand{\C}{{\mbb C}}
\newcommand{\CP}{{\C\mbb P}}
\newcommand{\Z}{{\mbb Z}}
\newcommand{\R}{{\mbb R}}

\begin{document}

\topmargin -2pt

\headheight 0pt

\topskip 0mm \addtolength{\baselineskip}{0.20\baselineskip}
\begin{flushright}
{\tt hep-th/0205186}
\end{flushright}

\vspace{10mm}

\begin{center}
{\large \bf   Deformed Hyperk\"{a}hler Structure for K3  Surfaces}\\

\vspace{12mm}

Chang-Yeong Lee\footnote{cylee@sejong.ac.kr}\\

{\it Department of Physics, Sejong University, Seoul 143-747, Korea}\\

\vspace{18mm}

\end{center}

\begin{center}
{\bf ABSTRACT}
\end{center}
We apply the method of algebraic deformation  to N-tuple of algebraic K3 surfaces.
When N=3, we show that 
the deformed triplet of algebraic K3 surfaces exhibits a
deformed hyperk\"{a}hler structure.
The deformation moduli space of this family of noncommutatively deformed K3 surfaces
turns out to be of dimension 57, which 
is three times of that of complex deformations of algebraic K3
surfaces. 
\\

\vfill


\thispagestyle{empty}

\newpage
\section*{I. Introduction}

Noncommutative geometry \cite{conn} is now an integral part of string/M theory \cite{sw}.
Since the work of Connes, Douglas, and Schwarz \cite{cds} 
connecting the noncommutative torus\cite{cr,rief} 
and the T-duality in the M theory context, various properties of noncommutative space
itself such as noncommutative tori and their varieties have been a subject of intensive
study \cite{nct,hv,t4ours,sw}. However, more interesting and complicated structures
such as noncommutative orbifolds and noncommutative 
 Calabi-Yau(CY) manifolds have been studied far less \cite{bl1,kl1,bs,ks1,ks2,kkl}.
Also, not much has been known about noncommutative spaces with complex structures.
Only recently, noncommutative tori with complex structures have been studied \cite{schwarz,manin,ds}.

In investigating the properties of noncommutative space with complex structure,
algebraic geometry approach seems to be a good fit.
In Ref.\cite{bjl}, Berenstein, Jejjala, and Leigh initiated an algebraic geometry approach
to noncommutative moduli space. Then applying this technique, 
 Berenstein and Leigh \cite{bl1} studied noncommutative CY threefolds;
a toroidal orbifold $T^6/{\Z}_2 \times
{\Z}_2$ and an orbifold of the quintic in ${\CP}^4$, each with
discrete torsion \cite{vafa,vfwt,doug,dgfl,jgomi}.
In their first example, 
they deformed the covering space in such a way that the center
of the deformed algebra
 corresponds to the commutative classical space, 
a CY threefold.
In that process, the complex structure of the center was also deformed 
as a consequence of the covering space deformation, 
and some part of the moduli space of complex deformations 
was indeed recovered.
They could also explain the fractionation of branes at singularities
 from noncommutative geometric viewpoint under the presence of
 discrete torsion.  
There, in order to be compatible with ${\Z}_2$ discrete torsion
 the three holomorphic coordinates $y_i$, the defining variables of
 the  three elliptic curves of  $T^6$,
became to anticommute with each other.

 In the commutative K3 case, the moduli space for the K3 space itself has been known already
 (see for instance \cite{aspinwall}),
and even the moduli space for the bundles on K3 surfaces
 has been studied \cite{mukai}.
In Ref. \cite{kl1},  algebraic deformation of  K3 surfaces has been studied 
 in the case of the orbifold $T^4/{\Z}_2$. 
There, the work was carried out by considering deformation of the invariants of the K3 itself, 
unlike the deformation of the variables of the covering space as in Ref. \cite{bl1}.

In Ref. \cite{kl2}, 
this method was applied for the algebraic K3  case.
Classically,
the complete family of  complex deformations of K3 surfaces is of 20 dimension
inside which that of the algebraic K3 surfaces is of 19 dimension \cite{aspinwall}.
In \cite{kl2},  a 19 dimensional
family of the noncommutative deformations of the 
general algebraic K3 surfaces was considered. The construction was
similar to 
the Connes-Lott's ``two-point space" construction of the standard model \cite{cl90}.
It was done by  deforming a pair of algebraic K3 surfaces and 
was called  ``two-point deformation". 
It was further generalized to the N-point case by considering
the deformation of N-tuple of algebraic K3 surfaces.
In the N-point deformation, the dimension of deformation moduli
turned out to be $19 N (N-1)/2 $ \cite{kl2}.

In this paper, we examine the N-point deformation method in the N=3 case.
Considering a 57 dimensional
family of noncommutatively deformed K3 surfaces, we 
show that the N=3 case corresponds to a
noncommutative deformation
of the hyperk\"{a}hler structure of K3 surfaces. 

In section II, we explain the method of  N-point deformation for
 the algebraic K3 surfaces in detail for N=2.
In section III, we show that for N=3 this family of noncommutatively deformed K3 surfaces 
exhibit  deformed hyperk\"{a}hler structures.
In section IV, we conclude with discussion.

\section*{II. 2-point deformation }\label{two-point}

In this section, we explain the method of N-point deformation \cite{kl2} of
algebraic K3 surfaces, specifically for the N=2 case
by considering the ``two-point space" version of noncommutative deformation for 
a pair of algebraic K3 surfaces. 

The N-point method was carried out by a direct extension 
of the algebraic deformation done for
the $T^4/{\Z}_2$ case \cite{kl1}.
General algebraic K3 surfaces are given by the following form
and with a point added at infinity.
\begin{equation}
\label{algk3}
y^2 = f(x_1, x_2)
\end{equation}
Here $f$ is a function with total degree 6 in $x_1,x_2$.

Now, we compare this with the Kummer surface,
the orbifold of $T^4/{\Z}_2$ case \cite{kl1}. 
 There $T^4$ was considered as the product of two elliptic curves, each given in the Weierstrass form
\begin{equation}
\label{t2}
y_i^2 = x_i (x_i-1)(x_i - a_i)
\end{equation}
with a point added at infinity for $i=1,2.$
By the following change of variables, the point at infinity is brought to a finite point:
\begin{eqnarray}
\label{t2tr}
y_i \lra y_i' = \frac{y_i}{x_i^2}, \\
x_i \lra x_i' = \frac{1}{x_i} . \nonumber
\end{eqnarray}


For algebraic K3 surfaces, we first consider a function with total degree 6
in complex variables $u,v,w,$ for instance
\[ F(u,v,w) =u^2v^3w + u^4v^2 . \]
In a patch where the point at infinity of $w$ can be brought to a finite point, 
dividing the both sides by $w^6$ the above expression can be rewritten as
\[ f(x_1,x_2) =x_1^2 x_2^3 +x_1^4 x_2^2  \]
where $x_1=\frac{u}{w}, x_2=\frac{v}{w} .$
The corresponding algebraic K3 surface is given by
\begin{equation}
\label{algk3p1}
y^2 = f(x_1, x_2) = x_1^2 x_2^3 +x_1^4 x_2^2  .
\end{equation}
Similarly, in a patch where the point at infinity of $u$ can be brought to a finite point,
we can reexpress it as
\begin{equation}
\label{algk3p2}
{y'}^2 = f'(x_1', x_2') = {x_1'}^3 x_2' + {x_1'}^2 
\end{equation}
where $x_1'=\frac{v}{u} = \frac{x_2}{x_1}, x_2' =\frac{w}{u} = \frac{1}{x_1} . $
Thus, in the case of the general algebraic K3, a point at infinity in one patch 
can be brought to a finite point in another patch by the following change of
variables 
\begin{eqnarray}
\label{k3ytr}
y \lra y' = \frac{y}{x_1^3}, \\
\label{k3xtr}
x_1 \lra x_1' = \frac{x_2}{x_1} , \\ 
x_2 \lra x_2' = \frac{1}{x_1} .
\nonumber
\end{eqnarray}

We now consider a noncommutative deformation of algebraic K3 surfaces.
 Following the same reasoning  as in Ref. \cite{kl1},
we consider two commuting complex variables $x_1, x_2$ and 
two noncommuting variables $t_1, t_2$ such that
\begin{eqnarray}
\label{k3ftn}
t_1^2 = h_1(x_1, x_2), \\
t_2^2 = h_2(x_1, x_2),
\nonumber
\end{eqnarray}
where $h_1, h_2$ are commuting functions of total degree 6 in $x_1, x_2$.
To be consistent with the condition that
 $t_1^2, t_2^2$ belong to the center, one can allow
the following deformation for $t_1, t_2$.
\begin{equation}
\label{Pctr}
t_1 t_2 + t_2 t_1 = P(x_1, x_2)
\end{equation}
Here the right hand side should be a polynomial and free of poles in each patch.
Thus, under the change of variables (\ref{k3xtr})
\begin{eqnarray*}
x_1 \lra x_1' = \frac{x_2}{x_1} , \\ 
x_2 \lra x_2' = \frac{1}{x_1}, 
\end{eqnarray*}
$t$'s should be changed into
\begin{equation}
\label{ttr}
t_i  \lra t_i' = \frac{t_i}{x_1^3},  \  \  \  \  {\rm for} \ \  i=1,2  . 
\end{equation}
This is due to the fact that $t$'s transform just like $y$ in (\ref{k3ytr}).
Therefore, $P$ transforms as
\begin{equation}
\label{Ptr}
P(x_1, x_2)  \lra  x_1^6 P' ( \frac{x_2}{x_1},  \frac{1}{x_1} ).  
\end{equation}
This implies that $P'$ should be of total degree 6 in $x_1', x_2'$, at most. 
Interchanging the role of $P$ and $P'$ one can see that $P$ should  be also
of total degree 6 in $x_1, x_2$. 

The above structure was understood as follows.
If the condition (\ref{Pctr}) is not imposed, then 
there exist two  independent commutative K3 surfaces. 
Once the condition (\ref{Pctr}) is imposed, these two
commutative K3 surfaces become a combined surface
in which the two K3 surfaces intertwined each other  
everywhere on their surfaces and becoming fuzzy. 
This seems to be similar to the two-point space version of the Connes-Lott model \cite{cl90}.
In the Connes-Lott model, every point of the space
becomes fuzzy due to the 1-to-2 correspondence at each point in the space,
where the two corresponding points at each classical location are fixed.
On the other hand, the present case is similar to
the relation between position $x$ and momentum $p$ 
in quantum mechanics at every point in the space.
However, since the two copies of the classical space
are combined to become a noncommutative space  just like the Connes-Lott
model, this construction was also called  
 two-point deformation
though its nature is a little different from the Connes-Lott's.

To count the dimension of the deformation moduli,
one simply needs to count the dimension of the
polynomials of degree 6 in three variables from (\ref{Ptr}) 
 up to constant modulo
projective linear transformations of three variables. 
Namely,
$28-1-8=19$, where 28 is the dimension of polynomials of degree 6 in
three variables and 1 and 8 correspond to a constant and
$PGL(3, \C)$, respectively.

\section*{III. Deformed hyperk\"{a}hler structure }\label{cp3}

In this section, we consider the N-point deformation for N=3.
 Following the method of the two-point deformation in the previous section, 
we consider commuting variables $x_1, x_2$ and three noncommuting variables $t_1, t_2, t_3$.
Here, each $t_i^2$ should belong to the center and be a function of 
total degree 6 in $x_1, x_2$, such that
\begin{eqnarray}
t_1^2 = h_1(x_1, x_2),  \nonumber \\
t_2^2 = h_2(x_1, x_2), \label{kfn} \\ 
t_3^2 = h_3(x_1, x_2),
\nonumber
\end{eqnarray}
 where $h_1, h_2, h_3$ are commuting functions of total degree 6 in $x_1, x_2$.
To be consistent with the condition that
 $t_i^2$ belong to the center, we can allow
the following deformation for $t_i$'s.
\begin{equation}
\label{kct}
t_i t_j + t_j t_i = P_{ij}(x_1, x_2),  ~~~~ i, j=1,2,3 ~( i \neq j).
\end{equation}
Here $P_{ij}$ should be polynomials and free of poles in each patch.
Thus, when we change from one patch to another, for instance
under the change of variables (\ref{k3xtr}) in the previous section;
\begin{eqnarray*}
x_1 \lra x_1' = \frac{x_2}{x_1} , \\ 
x_2 \lra x_2' = \frac{1}{x_1}, 
\end{eqnarray*}
$t_i$ should be changed into
\begin{equation}
\label{thtr3}
t_i  \lra t_i' = \frac{t_i}{x_1^3},  \  \  \  \  {\rm for} \ \  i=1,2,3  . 
\end{equation}
This is due to the fact that $t_i$ transform just like $y$ in (\ref{k3ytr})
in the previous section under the above change of patches.
Therefore, under the above change of variables $P_{ij}$ transform as
\begin{equation}
\label{Pijtr}
P_{ij} (x_1, x_2)  \lra  x_1^6 P_{ij}' ( \frac{x_2}{x_1},  \frac{1}{x_1} ).  
\end{equation}
By the same reasoning as in the two-point deformation case, one can see 
that each $P_{ij}$ is of total degree 6 in $x_1, x_2$, at most.
It is not difficult to show that one can also 
get the same conclusion for different changes of patches. 
Here, the condition (\ref{kfn}) for $t_i^2$ represents the 
different complex deformations of K3 surfaces, 
and its moduli space is of complex dimension 19. 
The condition (\ref{kct}) provides a characteristic of noncommutativity
for otherwise three commutative (algebraic) K3 surfaces
given by (\ref{kfn}).
Since each $P_{ij}$ is a polynomial of total degree 6 in $x_1, x_2$,
the  condition (\ref{kct})  makes the moduli space of
the above noncommutatively deformed K3 surfaces 
be of complex dimension 57.
 This is exactly
 three times of the moduli dimension of
complex deformations of the commutative algebraic K3 surfaces
that we mentioned above.
And it is different from the commutative hyperk\"{a}hler K3 case, in which the moduli space 
is of real dimension 58 as we will discuss below.
Then, what is the relationship between our newly constructed noncommutatively deformed
 K3 surfaces and the hyperk\"{a}hler structure of commutative K3 surfaces?

Before we address this question, we first review the property of
the moduli space $\cal{M}$ of Ricci flat metrics on a K3 surface $S$. 
If a given metric $g$ satisfies $g(Jv,Jw) =g(v,w)$ for any tangent vector $v,w$, then we 
say that the metric $g$ is compatible with the complex structure $J$.
If the two form $\Omega ( \cdot , \cdot ) = g (J \cdot , \cdot ) $ is closed,
then it is called a K\"{a}hler metric and $\Omega$ is called a K\"{a}hler form. 
Any given Ricci-flat metric $g$ induces a
Hodge $*$ operator on $H^2(S,\R) \cong {\R}^{3,19} $ by which $H^2(S,\R)$ 
can be decomposed as a direct sum of two eigenspaces, self dual part (eigenvalue 1)
of dimension 3 and anti-self dual part (eigenvalue $-1$) of dimension 19.
The self dual part is positive definite with the integration on $S$ after
wedge product, so that the moduli space
of Ricci-flat metrics is locally isomorphic to 
$ (O(3,19) / O(3) \times O(19) ) \times {\R}_+  $.
This is because $H^2(S,\R)$  has the intersection form (3,19) and 
the parameter of the scaling of the metric is ${\R}_+$. 
So the real dimension of  $\cal{M}$ is
$3 \times 19 + 1 = 58$.

We can also understand this in a different setting. 
Let ${\cal N} = \{ (J, \Omega ) \mid \Omega   $ is a  K\"{a}hler form
in the K3 surface with the complex structure  $  J \}$.
Then the real dimension of ${\cal N}$ is equal to the real dimension of the moduli space 
of complex structures plus the real dimension of K\"{a}hler forms, which is 
$  40 + 20 = 60 $.
We can define a map $\Phi$ from $\cal N$ to ${\cal M}$ as follows.
\[ \Phi ( ( J, \Omega ) ) = g, ~~~ {\rm such ~~  that} ~~~ g(\cdot , \cdot) =
\Omega( \cdot , J \cdot ) . \]
Then it is onto but not 1-to-1. The inverse image of $g$ by $\Phi$ is ${\mbb P}^1$.
So,  \[  {\rm dim}_{\R}{\cal M} = {\rm dim}_{\R}  {\cal N} -2  = 60 -2 = 58. \]

Now, we define the        
hyperk\"{a}hler structure on $S$.
In the first setting, for the given Ricci-flat metric $g$, the self dual part $\Lambda^+$
is a 3-dimensional real vector space consisting of vectors whose self intersection is positive.
To any compatible complex structure $J$ to $g$, we associate $\Omega$ which is a vector
in $\Lambda^+$ and is a (1,1) form. Then real (2,0) and (0,2) forms in $\Lambda^+$
are exactly the orthogonal to $\Omega$. 
Different compatible structrues $J$ to $g$ correspond to different unit vectors in $\Lambda^+$,
and they form $S^2$ isomorphic to ${\mbb P}^1$, inverse of $\Phi^{-1} (g)$.
Here we choose three orthogonal unit vectors $\Omega_1, \Omega_2, \Omega_3$ in $\Lambda^+$
such that the corresponding complex structures $J_1, J_2, J_3$ satisfy the relation
$J_i J_j = \epsilon_{ijk} J_k - \delta_{ij}$ for $i,j,k=1,2,3$.
This is called a hyperk\"{a}hler structure on $S$.

Now we return to our question of the connection 
between our N=3 construction and the hyperk\"{a}hler structure of K3 surfaces.
When $P_{ij}$ all vanish in (\ref{kct}), the $t_i ~ (i=1,2,3)$ in (\ref{kct}) satisfy the same relation as
the complex structures $J_i ~ (i=1,2,3)$ in
the case of the commutative hyperk\"{a}hler K3 surfaces, and
 $t_i$'s actually correspond to the
 complex structrues of the commutative K3 surfaces.
We see this as follows. 
If we consider just one of the $t_i$'s and disregard other two $t_i$'s for a moment,
then the $t_i$ represents a family of commutative algebraic K3 surfaces whose moduli 
dimension is of complex dimension 19.
On the other hand, when all the three $t_i$'s are present but
all the $P_{ij}$  vanish in (\ref{kct}),  then
the $t_i$'s are related like the $J_i$'s
of the commutative hyperk\"{a}hler K3 case.
However, when all $P_{ij}$ do not vanish and are independent of each other,
$t_i$'s become all independent and the moduli dimension becomes three times larger
than that of each piece represented by one of  the $t_i$'s.
Thus, the space becomes noncommutative under the condition (\ref{kct})
provided that all $P_{ij}$ do not vanish and are independent of each other.
Therefore, we can regard our new noncommutatively deformed K3 surfaces
having a deformed hyperk\"{a}hler structure for K3 surfaces,
since $t_i$'s satisfying  (\ref{kfn}) and (\ref{kct})  with  vanishing $P_{ij}$  
do admit the hyperk\"{a}hler structure \cite{aspinwall, frol1}.

This we can see by redefining $t_j ~ (j=1,2,3)$ as
$ t_j =i  \sqrt{h_j(x_1,x_2)} {\hat{t}}_j $ for $j=1,2,3.$
Then, in terms of $ {\hat{t}}_j, ~ $  (\ref{kfn}) and (\ref{kct}) become
\begin{eqnarray}
 {{\hat{t}}_1 }^2 = -1,  \nonumber \\
 {{\hat{t}}_2 }^2 = -1,  \label{kfnp} \\ 
 {{\hat{t}}_3 }^2 = -1,  
\nonumber
\end{eqnarray}
and 
\begin{equation}
\label{kctp}
{\hat{t}}_i {\hat{t}}_j  + {\hat{t}}_j {\hat{t}}_i   = - P_{ij}(x_1, x_2)/ \sqrt{h_i h_j},  ~~~~ i, j=1,2,3 ~( i \neq j).
\end{equation}
Recall that the quaternioc structure of the hyperk\"{a}hler structure of K3 surfaces
can be expressed as
\begin{eqnarray*}
 {J_i}^2  & = & -1,   \\
J_i J_j + J_j J_i  & = & 0,  ~~ i,j=1,2,3 (i \neq j).
\end{eqnarray*}
Comparing with this we  see that newly defined 
$ \{ \hat{t}_j  \}$ exhibit a deformed hyperk\"{a}hler structure 
for K3 surfaces.

The above construction of noncommutatively deformed hyperk\"{a}hler structure for
 K3 surfaces is a little different from
the one constructed in Ref.\cite{frol1,frol2} whose $T_i ~ (i=1,2,3)$ operators
do the similar role as our $t_i$'s. 
In Ref.\cite{frol1,frol2}, the commutation relation among $T_i$'s
was not deformed.
However, in their construction there exist extra anticommuting operators which provide 
holomorphic structures, and we wonder whether these additional anticommuting operators
could make the two constructions equivalent.


\section*{IV. Discussion}

In this paper, we deformed the hyperk\"{a}hler and complex structures
of K3 surfaces together. 
In the deformation of hyperk\"{a}hler structure, we introduced three noncommuting 
variables which correspond to three copies of commutative K3 surfaces and at the
same time represent three different complex structures of K3 surfaces.
Before deformation, we make these three variables have the same relation as the three complex
structures $J_i ~(i=1,2,3)$ of hyperk\"{a}hler K3 in which $J_i$'s posess 
the quaternionic structure 
and anticommute with each other.
Here, one may wonder that the deformation condition (\ref{kct}) could also be satisfied
with commuting variables when the polynomials $P_{ij}$ do not vanish.
That is possible, but the consequences are totally different 
depending on whether these variables are commuting or noncommuting ones.
When they are commuting variables, $P_{ij}$ in (\ref{kct}) are not independent
and they all can be expressed in terms of $h_i ~(i=1,2,3)$ functions in (\ref{kfn}).
Thus, there are only complex deformations and no noncommutative deformations.
On the other hand, when these variables are noncommuting ones
and $P_{ij}$ are nonvanishing, then
 $P_{ij}$ in (\ref{kct}) are all independent of $h_i $ functions.
Hence, we have both complex deformations from $h_i$ and noncommutative
deformations from $P_{ij}$. And since our construction is a deformation from the commutative
hyperk\"{a}hler structure of  K3 surfaces in the noncommutative direction, 
we end up with a noncommutatively deformed hyperk\"{a}hler
structure for K3 surfaces.

About the moduli dimension of  our noncommutatively deformed hyperk\"{a}hler structures
for K3, we still do not have a clear understanding of how  ours is related with the commutative one.
In the commutative case, it has real moduli dimension 58 as we explained before.
On the other hand, our noncommutatively deformed hyperk\"{a}hler structure
has complex moduli dimension 57.
Apparently, ours is exactly twice of that of the commutative one, 
once we disregard the parameter of overall scaling 
in the commutative case.
Thus,  if we complexify the metric moduli, then it seems that we can fill the gap.
In the construction of noncommutative hyperk\"{a}hler structure for K3 in Ref.\cite{frol1,frol2},
there exist a set of anticommuting operators providing the complexification.
However, we do not have the corresponding variables 
in our construction as we 
mentioned briefly at the end of the last section.
Since the commutation relation of the operators 
representing the complex structures in those works 
are not deformed unlike our construction, there is some possibility that
our noncommuting variables may also posess the property of 
these anticommuting operators
in Ref.\cite{frol1,frol2}.
We will leave this investigation for our future work.

\pagebreak

\vspace{5mm}
\noindent
{\Large \bf Acknowledgments}

\vspace{5mm}
This work was supported by
KOSEF Interdisciplinary Research Grant No. R01-2000-00022.
We would like to thank Hoil Kim for many helpful discussions.


\newcommand{\MPL}{Mod.\ Phys.\ Lett.}
\newcommand{\NP}{Nucl.\ Phys.}
\newcommand{\PL}{Phys.\ Lett.}
\newcommand{\PR}{Phys.\ Rev.}
\newcommand{\PRL}{Phys.\ Rev.\ Lett.}
\newcommand{\CMP}{Commun.\ Math.\ Phys.}
\newcommand{\JMP}{J.\ Math.\ Phys.}
\newcommand{\JHEP}{JHEP}
\newcommand{\ib}{{\it ibid.}}


\end{document}